\documentclass{elsart}
\usepackage{graphics}
\usepackage{graphicx}
\usepackage{epsfig}
\usepackage{amssymb}

\begin{document}
\begin{frontmatter}
\title{Promotion of cooperation induced by the interplay between structure and game dynamics}
\author[ad1,ad2]{Feng Fu},
\ead{fufeng@pku.edu.cn}
\author[ad1,ad2]{Xiaojie Chen},
\author[ad1,ad2]{Lianghuan Liu},
\author[ad1,ad2]{Long Wang\corauthref{cor1}}
\ead{longwang@pku.edu.cn}

\corauth[cor1]{Corresponding author. Fax: +86-010-62754388.}
\address[ad1]{Center for Systems and Control, College of Engineering,
Peking University, Beijing 100871, China}
\address[ad2]{Department of Industrial Engineering and Management, College of Engineering, Peking University, Beijing 100871, China}

\begin{abstract}
We consider the coupled dynamics of the adaption of network
structure and the evolution of strategies played by individuals
occupying the network vertices. We propose a computational model
in which each agent plays a $n$-round Prisoner's Dilemma game with
its immediate neighbors, after that, based upon self-interest,
partial individuals may punish their defective neighbors by
dismissing the social tie to the one who defects the most times,
meanwhile seek for a new partner at random from the neighbors of
the punished agent. It is found that the promotion of cooperation
is attributed to the entangled evolution of individual strategy
and network structure. Moreover, we show that the emerging social
networks exhibit high heterogeneity and disassortative mixing
pattern. For a given average connectivity of the population and
the number of rounds, there is a critical value for the fraction
of individuals adapting their social interactions, above which
cooperators wipe out defectors. Besides, the effects of the
average degree, the number of rounds, and the intensity of
selection are investigated by extensive numerical simulations. Our
results to some extent reflect the underlying mechanism promoting
cooperation.
\end{abstract}

\end{frontmatter}

\section{Introduction}
Cooperative behaviors are ubiquitous in real world, ranging from
biological systems to socioeconomic systems. However, the question
of how natural selection can lead to cooperation has fascinated
evolutionary biologists for several decades. Fortunately, together
with classic game theory, evolutionary game theory provides a
systematic and convenient framework for understanding the
emergence and maintenance of cooperative behaviors among selfish
individuals~\cite{Neumann,Marnard}. Especially, the Prisoner's
Dilemma game (PDG) as a general metaphor for studying the
evolution of cooperation has attracted considerable
interests~\cite{Axelrod}.

In the original PDG, two players simultaneously decide whether to
cooperate (C) or to defect (D). They both receive $R$ upon mutual
cooperation and $P$ upon mutual defection. A defector exploiting a
C player gets $T$, and the exploited cooperator receives $S$, such
that $T>R>P>S$ and $2R>T+S$. As a result, it is best to defect
regardless of the co-player's decision. Thus, in well-mixed
infinite populations, defection is the evolutionarily stable
strategy (ESS), even though all individuals would be better off if
they cooperated. Thereby this creates the social dilemma, because
when everybody defects, the mean population payoff is lower than
that when everybody cooperates. In a recent review Nowak suggested
five rules for the evolution of cooperation (see
Ref.~\cite{Nowak06} and references therein). Most noteworthy,
departure from the well-mixed population scenario, the rule
``network reciprocity'' conditions the emergence of cooperation
among players occupying the network vertices~\cite{Ohtsuki06}.
That is, the benefit-to-cost ratio must exceed the average number
of neighbors per individual. Actually, the successful development
of network science provides a convenient framework to describe the
population structure on which the evolution of cooperation is
studied. The vertices represent players, while the edges denote
links between players in terms of game dynamical interactions.
Furthermore, interactions in real-world network of contacts are
heterogeneous, often associated with scale-free (power-law)
dependence on the degree distribution, $P(k)\sim k^{-\gamma}$ with
$2<\gamma<3$. Accordingly, the evolution of cooperation on model
networks with features such as
lattices~\cite{Szabo98,Szabo02,Szabo05,Vukov06},
small-world~\cite{Abra2001,Masuda03,Toma2006},
scale-free~\cite{Santos05,Santos2006a,Santos2006b}, and community
structure~\cite{Chen06} has been scrutinized. Interestingly,
Santos {\it et al.} found that scale-free networks provide a
unifying framework for the emergency of
cooperation~\cite{Santos05}.

From the best of our knowledge, so far much previous works of
games on networks are based on crystalized (static) networks, i.e.
the social networks on which the evolution of cooperation is
studied are fixed from the outset and not affected by evolutionary
dynamics on top of them. However, interaction networks in real
world are continuously evolving ones, rather than static graphs.
Indeed, individuals have adaptations on the number, frequency, and
duration of their social ties base upon some certain feedback
mechanisms. Instead of investigating the evolutionary games on
static networks which constitute just one snapshot of the real
evolving ones, recently, some researchers proposed that the
network structure may co-evolve with the evolutionary game
dynamics~\cite{Ebel02,Zimmermann04,Zimmermann05,Eguluz05,Renjie06,Santos2006c,Pacheco_PRL}.
Interestingly, as pointed out in
Refs.~\cite{Zimmermann04,Zimmermann05,Santos2006c}, the entangled
evolution of individual strategy and network structure constitutes
a key mechanism for the sustainability of cooperation in social
networks. Therefore, to understand the emergence of cooperative
behavior in realistic situations (networks), one should combine
strategy evolution with topological evolution. From this
perspective, we propose a computational model in which both the
adaptation of underlying network of interactions and the evolution
of behavioral strategy are taken into account simultaneously. In
our model, each agent plays a $n$-round Prisoner's Dilemma game
with its immediate neighbors, after that, based upon
self-interest, partial individuals may punish their defective
neighbors by dismissing the social tie to the one who defects the
most times, meanwhile seek for a new partner at random from the
neighbors of the punished agent. We shall show that such
individual's local adaptive interactions lead to the situation
where cooperators become evolutionarily competitive due to the
preference of assortative mixing between cooperators. The
remainder of this paper is organized as follows. In the following
section, the model is introduced in detail. Sec.~III presents the
simulation results and discussions. We finally draw conclusions in
Sec.~IV.

\section{The model}
We consider a symmetric two-player game where $N$ individuals
engage in the Prisoner's Dilemma game (PDG) over a network. The
total number of edges $M$ is fixed during the evolutionary
process. Each individual $i$ plays with its immediate neighbors
defined by the underlying network. The neighbor set of individual
$i$ is denoted as $\Omega_i$, which is allowed to evolve according
to the game results. Let us denote by $s_i$ the strategy of
individual $i$. Player $i$ can follow two simple strategies:
cooperation [C, $s_i=(1,0)^T$] and defection [D, $s_i=(0,1)^T$] in
each round. Following previous studies~\cite{Nowak92,Nowak93}, the
payoff matrix $M$ has a rescaled form depending on a single
parameter,
\begin{equation}
M=\left (\begin{array}{cc}
  1 \,& \,0 \\
  b \,& \,0 \\
\end{array}\right ),
\end{equation}
where $1<b<2$.\\
In each round, each agent plays the same strategy with all its
neighbors, and accumulates the payoff, observing the aggregate
payoff and strategy of its neighbors. The total income of the
player at the site $x$ can be expressed as
\begin{equation}
P_x=\sum_{y\in \Omega_x}s_x^TMs_y,
\end{equation}
where the sum runs over all the neighboring sites of $x$,
$\Omega_x$. In evolutionary games the players are allowed to adopt
the strategies of their neighbors after each round. Then, the
individual $x$ randomly selects a neighbor $y$ for possibly
updating its strategy. The site $x$ will adopt $y$'s strategy with
probability determined by the total payoff difference between
them~\cite{Szabo98,Blume93,Traulsen06}:
\begin{equation}
\label{transp} W_{s_x \leftarrow s_y} =
\frac{1}{1+\exp[\beta(P_x-P_y)]},
\end{equation}
where the parameter $\beta$ is an inverse temperature in
statistical physics, the value of which characterizes the
intensity of selection. $\beta\to0$ leads to neutral (random)
drift whereas $\beta\to\infty$ corresponds to the imitation
dynamics where the $y$'s strategy replaces $x$'s whenever
$P_y>P_x$. For finite value of $\beta$, the larger $\beta$ is, the
fitter strategy is more apt to replace to the less one, thus the
value of $\beta$ indicates the intensity of selection.

\begin{figure}
\centering
\includegraphics[width=6cm]{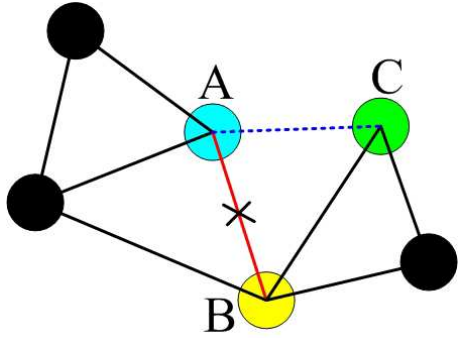}
\caption{Illustration of individual's local adaptive interactions.
Assuming A is picked for updating its social ties after playing
$n$-round Prisoner's Dilemma game with immediate neighbors. A
dismisses the link to B, who defects the most times, and rewires
the link to C, a random neighbor of B. \label{fig1}}
\end{figure}

In the present model, we assume each agent plays a $n$-round PDG
with its neighbors ($n\geq1$), and then $m$ randomly selected
individuals are allowed to adapt their social ties according to
the game results ($1 \leq m \leq N$). Here the individuals are
endowed with the limited cognitive capacities---each agent records
the strategies of its opponents used in the $n$-round game. Then
they are able to decide to maintain those ties from which they
benefit from, and to rewire the adverse links. For the sake of
simplicity, if someone is picked for updating its neighbors, only
the most disadvantageous edge is rewired. It dismisses the link to
the one, who defects the most times (if there exist more than one
individuals who defect the same maximum times, the one is chosen
at random), and redirects the link to a random neighbor of the
punished (see Fig.~\ref{fig1} as an illustrative example). The
advantage of rewiring to neighbor's neighbor is twofold: first,
individuals tend to interact with others that are close by in a
social manner~\cite{Watts06}, i.e. friend's friend is more likely
to become a friend (partner); second, every agent is seeking to
attach to cooperators, thus redirecting to neighbor's neighbor
will be a good choice since the neighbor also tries to establish
links with cooperators. Hence rewiring to a neighbor of a defector
is no doubt a good choice for individuals with local information
only~\cite{Santos2006c}. Herein, the parameters $n$ and $m$ can be
viewed as the corresponding time scales of strategy evolution and
network structure adaptation. As our strategy evolution uses
synchronous updating, while evolution of network topology adopts
asynchronous updating, in our case, the strategy updating event
proceeds naturally much more frequent than evolution of network
structure (as $N\cdot n>m$). Nevertheless, even though network
structure adaption is much lower than game dynamics, cooperation
is still promoted by the efficient interplay between the two
dynamics.

Let us point out the differences between our model and previous
works. In Refs.~\cite{Zimmermann04,Zimmermann05}, the evolution of
strategy adopted the ``Best-Take-Over'' update rule where each
agent imitates the strategy of the best neighbor. Besides,
individuals are divided into two types based on the payoffs:
satisfied and unsatisfied. If individual's payoff is the highest
among its neighbors, then it is satisfied. Otherwise, it is
unsatisfied. The network adaptation dynamics is restricted to the
ones who are defectors and unsatisfied. Thus the unsatisfied
defector breaks the link with defective neighbor with probability
$p$, replaces it by randomly choosing agent uniformly from the
network. More recently, Ref.~\cite{Santos2006c} proposed another
minimal model that combined strategy evolution with topological
evolution. They used asynchronous update rule both for evolution
of strategy and structure by the Fermi function. In their model,
topological evolution is manipulated in the following way: a pair
of C-D or D-D are chosen at random, the one may compete with the
other to rewire the link, rewiring being attempted to a random
neighbor's neighbor with certain probability determined by payoff
difference between them. Whereas in our model, we argue that
individuals are exclusively based on their self-interest. Even if
the individual is a cooperator, it could not bear the exploitation
by defectors. Furthermore, in our situation, individuals have
enough inspection over their opponents because they are engaged in
a $n$-round PDG. Subsequently, each agent can punish the most
defective neighbor by dismissing the link, meanwhile seeks to
establish links with cooperators. Especially, in our model the
agents are endowed with limited memory abilities by which they can
punish the most defective one. In addition, the way of associated
time scales with respect to evolution of strategy and structure is
different in our model from the previous related works. As
aforementioned, after playing $n$-round PDG with neighbors,
individuals update their local interactions according to the game
results. Such timely feedback mechanism is ubiquitous in natural
world. Besides, the adaption of network structure is much slower
than evolution of strategies in our model. Such feature reflects
the fact that individuals may not respond rapidly to the
surrounding since maintaining and rewiring interactions are costly
to them. In previous investigations, the time scale often is
implemented in a stochastic manner. Although the implementation of
time scales in the literature is equivalent to our model in a way,
our method may be more plausible. Therefore, our model is
different from the previous ones in these respects and captures
the characteristics in real situation. In what follows, we
investigate under which conditions cooperation may thrive by
extensive numerical simulations. And also, we show the effects of
the changing parameters in our model on the evolution of
cooperation.

\section{Simulation results and discussions}

We consider $N$ individuals occupying the network vertices. Each
interaction between two agents is represented by an undirected
edge (a total of $N_E$). The social networks are evolving in time
as individuals adapt their ties. The average connectivity $\langle
k \rangle=2N_E/N$ is conserved during the topological evolution
since we do not introduce or destroy links. This point assumes a
constrained resource environment, resulting in limited
possibilities of network configurations. Besides, we impose that
nodes linked by a single edge can not lose this connection, thus
the evolving networks are connected at all times. We calculated
the amount of heterogeneity of the networks as
$h=N^{-1}\sum_kk^2N(k)-\langle k \rangle^2$ (the variance of the
network degree sequcence), where $N(k)$ gives the number of
vertices with $k$ edges. Additionally, in order to investigate the
degree-degree correlation pattern about the emerging social
networks, we adopted the assortativity coefficient $r$ suggested
by Newman~\cite{degree-cor},
\begin{equation}
r=\frac{M^{-1}\sum_ij_ik_i-[M^{-1}\sum_i\frac{1}{2}(j_i+k_i)]^2}{M^{-1}\sum_i\frac{1}{2}(j_i^2+k_i^2)-[M^{-1}\sum_i\frac{1}{2}(j_i+k_i)]^2},
\end{equation}
here $j_i,k_i$ are the degrees of the vertices at the ends of the
$i$th edge, with $i=1,\cdots,N_E$. Networks with assortative
mixing pattern, i.e. $r>0$, are those in which nodes with large
degree tend to be connected to other nodes with many connections
and vice versa.

The interplay between network structure and game dynamics is
implemented as following steps:
\begin{itemize}
    \item Step (1): The evolution of strategy uses synchronous updating. Each
    agent plays the PDG with its immediate neighbors for consecutive $n$
    rounds. After each round, each individual adapts its strategy
    according to Eq.~(\ref{transp}), and records down the defection times of
    its every neighbors.
    \item Step (2): The update of individual's local social
    interactions is asynchronous. $m$ agents are successively chosen at random
    to rewire the most adverse links (if any) as shown in Fig.~\ref{fig1}.
    \item Step (3): Repeat the above two steps until the population converges to an absorbing
    state (full cooperators or defectors), or stop repeating the above two steps after $10^5$
    generations.
\end{itemize}

We start from a homogeneous random graph by using the method in
Ref.~\cite{Santos05b}, where all nodes have the same number of
edges, randomly linked to arbitrary nodes. Initially, an equal
percentage of cooperators and defectors is randomly distributed
among the elements of the population. We run 100 independent
simulations for the corresponding parameters $N$, $\langle k
\rangle$, $n$, $m$, $b$, and $\beta$. We also compute the fraction
of runs that ended up with 100\% cooperators. If the evolution has
not reached an absorbing state after $10^5$ generations, we take
the average fraction of cooperators in the population as the final
result. Moreover, we observe the time evolution of the network
structure and strategy, including the degree-degree correlation
coefficient, the degree of heterogeneity, the frequency of
cooperators, the fraction of C-C/C-D/D-D links, etc. Finally, we
confirm that our results are valid for different population size
$N$ and edge number $N_E$.

\begin{figure}
\centering
\includegraphics[width=10cm]{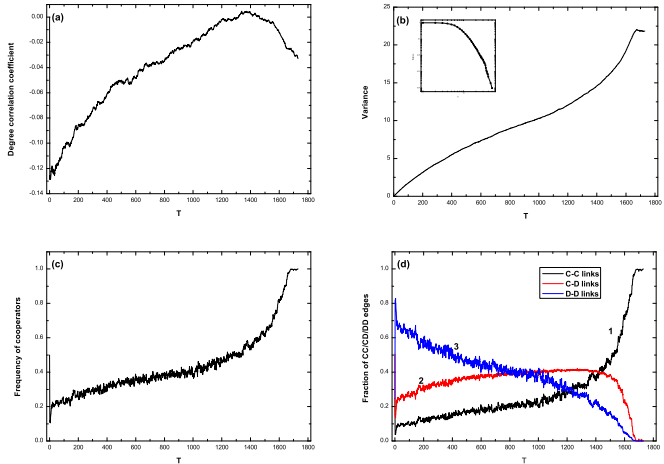}
\caption{Time evolution of the network structure as a result of
the adaption of social ties: (a) the degree correlation
coefficient, (b) the amount of heterogeneity, measured in terms of
the variance of the degree distribution, (c) the frequency of
cooperators, and (d) the fraction of C-C/C-D/D-D edges. The inset
in panel (b) shows the cumulative degree distribution in steady
state. The network evolution starts from a homogeneous random
graph, in which all nodes have the same number of edges ($\langle
k \rangle$), randomly linked to arbitrary nodes. The corresponding
parameters are $N=10^4$, $\langle k \rangle =8$, $b=1.2$, $n=6$,
$\beta=50$, and $m=100$.\label{fig2} }
\end{figure}

We report a typical time evolution of the network structure as a
result of the adaption of social ties in Fig.~\ref{fig2}, with
relevant parameters $N=10^4$, $\langle k \rangle =8$, $b=1.2$,
$n=6$, $\beta=50$, and $m=100$. The emerging social network shows
disassortative mixing pattern, indicating that large-degree nodes
tend to be connected to low-degree nodes. The degree-degree
correlation coefficient $r$ of the network we started from is
zero. Once the network structure adaption is in effect,
disassortative mixing pattern will be developed. Since the
rewiring process is attempted to a random neighbor's neighbor,
thus the nodes with large connectivity are more possible to be
attached by others. Due to such ``rich gets richer'',
inhomogeneity is induced as shown in Fig.~\ref{fig2}(b). The
amount of the heterogeneity (degree variance) $h$ increases in
virtue of the rewiring process. The inset in Fig.~\ref{fig2}(b)
plots the cumulative degree distribution of the stationary (final)
network, which exhibits high heterogeneity with a power-law tail.
Fig.~\ref{fig2}(c) displays the evolution of cooperation. We find
that the frequency of cooperators decreases at first due to the
temptation to defect, and then because of the adaptive
interactions, the cooperation level thrives gradually and the
population converges into an absorbing state of 100\% cooperators.
The viability of cooperation is also in part promoted by the
heterogeneity of the underlying heterogeneity. From
Fig.~\ref{fig2}(d), we can see that local assortative interactions
between cooperators is enhanced by structural updating, while
assortative interactions between defectors and defectors is
inhibited remarkably. The disassortativity between cooperators and
defectors is promoted in the beginning by strategy updating,
however, being diminished eventually by structural updating.
Clearly, It is thus indicated that the interplay between strategy
and structure will facilitate the emergence of cooperation.

\begin{figure}
\centering
\includegraphics[width=10cm]{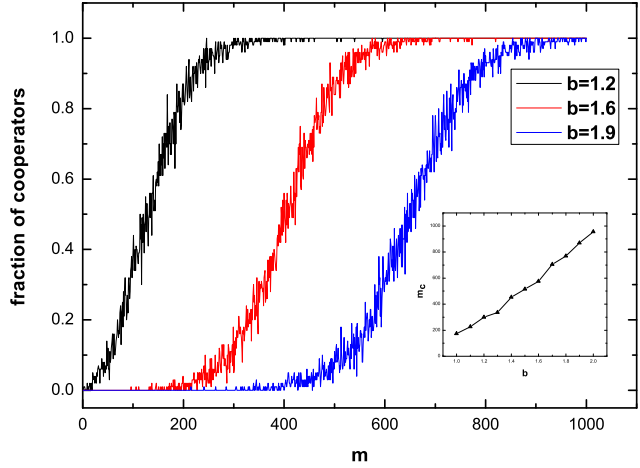}
\caption{Fraction of cooperators at end as a function of $m$ for
different values of $b$. We ran 100 simulations, starting from
$50\%$ cooperators. The values plotted correspond to the fraction
of runs which ended with $100\%$ cooperators. From left to right,
$b=1.2, 1.6, 1.9$ respectively. The inset plots the critical value
$m_c$ vs $b$. $N=10^3$, $\langle k \rangle =4$, $n=6$, and
$\beta=0.01$.\label{fig3}}
\end{figure}

\begin{figure}
\centering
\includegraphics[width=10cm]{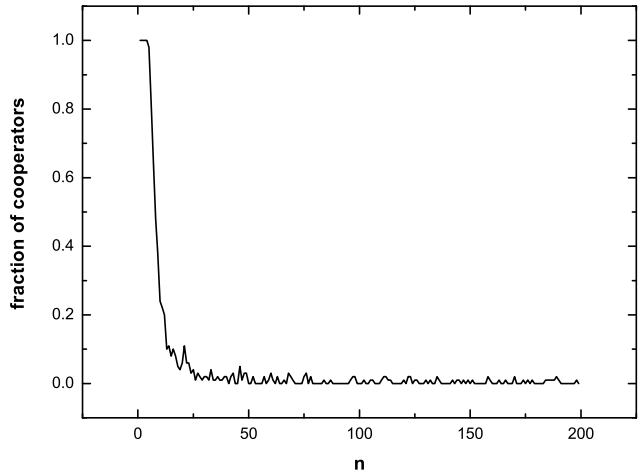}
\caption{The role of different average number of rounds $n$ to
evolution of cooperation for fixed $m=200$, $N=10^3$, $\langle k
\rangle =4$, $b=1.2$, and $\beta=0.01$.\label{fig4}}
\end{figure}

\begin{figure}
\centering
\includegraphics[width=10cm]{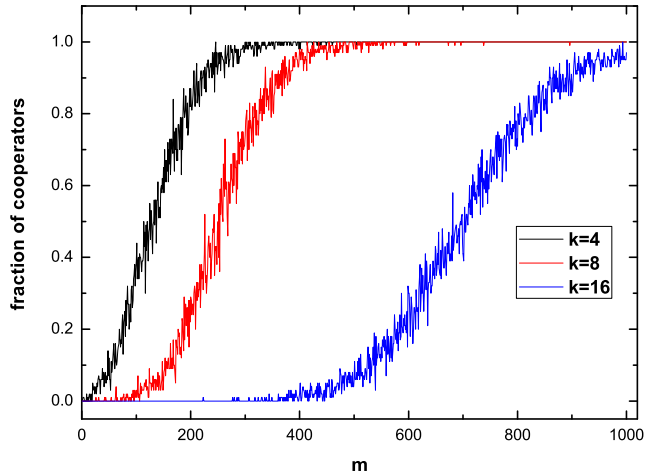}
\caption{The influence of average connectivity $\langle k \rangle$
to evolution of cooperation. Fraction of cooperators as a function
of $m$ for different values $\langle k \rangle$. From left to
right, $\langle k \rangle=4, 8, 16$ respectively. $N=10^3$, $n=6$,
$b=1.2$, and $\beta=0.01$.\label{fig5}}
\end{figure}

\begin{figure}
\centering
\includegraphics[width=10cm]{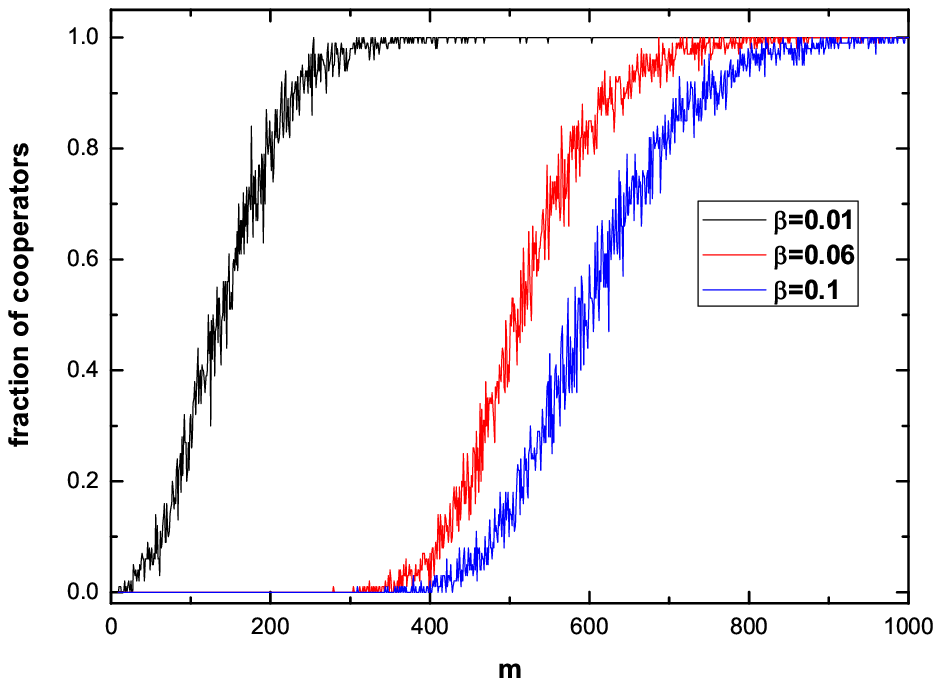}
\caption{Fraction of cooperators as a function of $m$ for
different values $\beta$. From left to right, $\beta=0.01, 0.06,
0.1$ respectively. $N=10^3$, $n=6$, $b=1.2$, and $\langle k
\rangle=4$.\label{fig6}}
\end{figure}

Let us consider the effect of the amount of temptation to defect
$b$ to the evolution of cooperation. The relevant result is
presented in Fig.~\ref{fig3}. With increasing $b$, the structural
updating event must be sufficiently frequent to guarantee the
survival of cooperators. In other words, the fraction of
individuals chosen for updating social ties should be accordingly
increased to ensure the sustainability of cooperators. When the
value of $b$ is enlarged, the defectors become more favorable by
the nature selection. Nevertheless, with the aid of structural
updating, a small fraction of surviving cooperators promotes them
into hubs (large degree nodes), since they are attractive to
neighborhood. Such co-evolution of strategy and structure leads to
highly heterogeneous networks in which the cooperators become
evolutionarily competitive as demonstrated in
Refs.~\cite{Santos05,Santos2006a,Santos2006b}. For fixed $b$, we
observed a critical value $m_c$ for $m$, above which the
cooperator will wipe out defectors. For fixed number of rounds
$n$, the critical value $m_c$ monotonously increases with
increasing $b$, as shown at the inset of Fig.~\ref{fig3}.
Therefore the prompt network adaption prevents cooperators from
becoming extinct, further, resulting in an underlying
heterogeneous social network which is the ``green house'' for
cooperators to prevail under strategy dynamics. Consequently, the
entangled co-evolution of strategy and structure promotes the
evolution of cooperation among selfish individuals.

Furthermore, we investigated the effect of number of rounds $n$ to
the emergence of cooperation. Fix fixed $m$ and other parameters,
there exists a critical value for $n$, above which the cooperators
will vanish as shown in Fig.~\ref{fig4}. Indeed, although the
structural updating promotes the cooperators to a certain extent,
its role will be suppressed by the long-time strategy dynamics
(corresponding to large $n$). In our case, strategy dynamics is
synchronous while structural updating is asynchronous, namely, for
each repetition in the simulations, strategy updating happens at a
frequency of $N\cdot n$ while the structural updating occurs at a
frequency of $m$. Hence the evolution of strategy is much more
frequent than that of structure. Thus, with large $n$, able
defectors outperform cooperators through strategy dynamics, even
though the heterogeneity, resulting from structural updating, is
positive to evolution of cooperation. This result illustrates that
even if the evolution of network topology is less frequent than
the evolution of strategy, cooperators still have chances to beat
defectors under appropriate conditions.

As is well known, cooperation is promoted in the situation where
individuals are constrained to interact with few others along the
edges of networks with low average
connectivity~\cite{Ohtsuki06,Santos2006a,Nowak92}. To understand
the cooperation in real-world interaction networks of which the
average connectivity is normally relatively high, one needs new
insight into the underlying mechanism promoting cooperation. Here,
the role of average connectivity to evolution of cooperation is
inspected. In Fig.~\ref{fig5}, it is shown that for increasing
$\langle k \rangle$, the individuals must be able to promptly
adjust their social ties for cooperation to thrive, corresponding
to increasing $m$. Thus in order to explain the cooperation in
communities with a high average number of social ties, the
entangled co-evolution of network structure and strategy should be
taken into account simultaneously. On static networks, maximum
cooperation level occurs at intermediate average
degree~\cite{wwx}. Moreover, when the connections are dense among
individuals (large average connectivity $\langle k \rangle$),
cooperators die out due to mean-field behavior. Conversely, our
results suggest that even in highly-connected network, on account
of the proposed structural adaption, cooperators can beat back the
defectors and dominate the populations.

Finally, we report the influence of changing intensity of
selection $\beta$ on the evolution of cooperation in
Fig.~\ref{fig6}. It is indicated that reducing $\beta$ will demote
the influence of the game dynamics, thereby increase the
survivability of the less fit. Clearly, the smaller the value of
$\beta$ is, the smaller the critical value of $m$. In fact, for
small $m$, cooperators' survival probability increases with
decreasing $\beta$ although cooperators are generally less fit.
Such increased survivability enhances assortative interactions
between cooperators through network structure adaption. As a
result, the critical value of $m$ above which cooperators dominate
defectors decreases with decreasing $\beta$.

\section{Conclusions}
In summary, we have studied the coupled dynamics of strategy
evolution and the underlying network structure adaption. We
provided a computational model in which individuals are endowed
with limited cognitive abilities in the $n$-round PDG---limited
memories for recording the defection times of opponents. After the
$n$-round game, $m$ randomly chosen individuals are allowed to
adjust their social ties based on the game results. The values of
$n$ and $m$ are corresponding to the associated time scales of
strategy dynamics and structural updating respectively. We found
that for a given average connectivity of the population and the
number of rounds, there is a critical value for the fraction of
individuals adapting their social interactions above which
cooperators wipe out defectors. In addition, the critical value of
$m$ above which cooperators dominate defectors decreases with
decreasing intensity of selection $\beta$. Moreover, for
increasing average connectivity, the individuals must be able to
swiftly adjust their social ties for cooperators to thrive.
Finally, the emerging social networks at steady states exhibit
nontrivial heterogeneity which is the catalyst for emergence of
cooperation among selfish agents. To a certain extent, our results
shed some light on the underlying mechanism promoting cooperation
among selfish individuals, and also provide an alternative insight
into the properties accruing to those networked systems and
organizations in natural world.

\section*{Acknowledgement}
Delightful discussion with Dr. Wenxu Wang is gratefully
acknowledged. This work was supported by NNSFC (60674050 and
60528007), National 973 Program (2002CB312200), National 863
Program (2006AA04Z258) and 11-5 project (A2120061303).

\end{document}